\documentstyle[12pt]{article}

\def\balpha{\mbox{\boldmath$\alpha$}}
\def\bnabla{\mbox{\boldmath$\nabla$}}

\def\bsigma{\mbox{\boldmath$\sigma$}}
\def\bgamma{\mbox{\boldmath$\gamma$}}

\def\bk{{\bf k}}
\def\bp{{\bf p}}
\def\br{{\bf r}}
\def\bv{{\bf v}}
\def\bz{{\bf z}}
\def\bx{{\bf x}}

\def\bL{{\bf L}}

\def\bP{{\bf P}}
\def\bJ{{\bf J}}
\def\bA{{\bf A}}

\newcommand{\beq}{\begin{equation}}
\newcommand{\eeq}{\end{equation}}
\newcommand{\bea}{\begin{eqnarray}}
\newcommand{\eea}{\end{eqnarray}}
\newcommand{\rar}{\rightarrow}

\newcommand{\lan}{\langle}
\newcommand{\ran}{\rangle}

\newcommand{\p}{\partial}
\begin{document}

\font\fortssbx=cmssbx10 scaled \magstep2
\hbox to \hsize{
\includegraphics{uwlogo.ps}
\hskip.5in \raise.1in\hbox{\fortssbx University of Wisconsin - Madison}
\hfill$\vcenter{\hbox{\bf MADPH-95-885}
            \hbox{May 1995}}$ }
\vskip 2cm
\begin{center}
\Large
{\bf Flux Tubes in Effective Field Theory} \\
\vskip 0.5cm
\large
M. G. Olsson and  Sini\v{s}a Veseli \\
\vskip 0.1cm
{\small \em Department of Physics, University of Wisconsin, Madison,
	\rm WI 53706} \\
\vspace*{+0.2cm}
 Ken Williams \\
{\small \em Continuous Electron Beam Accelerator Facility \\
	Newport News, VA 29606, USA \\
	and \\
\vspace*{-0.2cm}
Physics Department, Hampton University, Hampton, VA 29668}
\end{center}
\thispagestyle{empty}
\vskip 0.7cm

\begin{abstract}
Quark-antiquark bound states are examined in the long-range strong-coupling
limit with the minimal area law of lattice gauge theory assumed as input.
Matrix element relations are established which in the effective theory obtain
dynamical equations equivalent to a formulation of the flux-tube model.
\end{abstract}

\newpage
\section{Introduction}

The success of QCD as the correct theory of strong interactions has
rested primarily upon perturbative calculations.
The prediction of hadronic properties
 is inherently non-perturbative
and has been much slower to develop. Most off lattice progress
has turned out to be
 largely irrelevant to  hadron dynamics. An exception has been
 the  low quark velocity
Wilson loop expansion \cite{eichten}. The aim of this paper is to
extend this method
to arbitrary quark velocity. Our principal
result
is that under natural assumptions QCD
provides a framework for meson dynamics that is
identical  with the relativistic flux tube (RFT) model \cite{low,ovw}.

The RFT model assumes
 that a chromoelectric flux tube
stretches in a straight line between the quark
and the antiquark in a meson. The tube rest frame
energy per unit length $a$ contributes
to rotational energy, momentum, and angular momentum
in the meson rest frame. Also, as pointed out by
Buchm$\ddot{\rm u}$ller \cite{buchmuller}, a pure chromoelectric
field in the  tube rest frame implies no long range spin-spin
correlation. In the last few years the RFT model has been demonstrated
to be numerically calculable for any quark mass
case \cite{numerical}. The RFT model provides intuitive
physical pictures for relativistic
corrections in agreement with QCD and becomes a Nambu-Goto
string for high rotational states
\cite{low}.

The QCD Lagrangian density is a fundamental object in the theory of strongly
interacting particles. For a meson it is given by the minimally transformed
free Lagrangian
\bea
p_{\mu}  &\rar & p_{\mu} - g A_{\mu} \ , \\
{\cal L}_{free} &\rar & {\cal L }_{QCD} = \sum_{j} \bar{q}_j  (\imath
\not\hspace{-1.25mm}{\it D} - m_j) q_j -\frac{1}{4}
F^a_{\mu \nu} F_a^{\mu \nu} \ , \label{lagrangian}
\eea
from which follow the Euler-Lagrange equations
\bea
(\imath \not\hspace{-1.25mm}{\it D} - m_j) q_j &=& 0 \ ,  \label{el1} \\
D^{\mu} F_{\mu \nu} &=& g \, \sum_j \bar{q}_j T \gamma_{\nu} q_j \ .
\label{el2}
\eea
These are non-linear and amenable only as a perturbation expansion in small
coupling $g$. Wilson's  observation \cite{wilson} over two decades ago has led
to some of the most fruitful work in the non-perturbative regime. The simple
picture offered in his area law for static quarks is compelling: Contributions
to the $ q \bar{q} $ propagator fall off exponentially in the area swept out by
lines of gauge invariant chromoelectric flux
joining the constituents's world lines; in this way, widely separated paths are
suppressed. Subsequent lattice simulations verify the same qualitative behavior
for dynamical quarks.

Progress in obtaining a consistent description of confinement for realistic
off-lattice calculations  has been largely uncertain. The variety of
semi-relativised confinement models based on the above simple idea is striking
and is partly due to the limited theoretical input available beyond the static
limit. Notable exceptions are in works of Eichten and Feinberg, Gromes and
later those of Brambilla and Prosperi \cite{eichten} where analytic expressions
are derived from the minimal area law to first order in  the inverse heavy
quark mass.
More recently, a relativistic QCD-string Lagrangian for spinless quarks
 \cite{dubin} has been deduced.

Here we derive an effective Hamiltonian for a quark-antiquark system in the
confinement region with the minimal area
(MA) law of lattice gauge theory as input. In this arrangement, the gauge
field's non-Abelian character is manifest through the minimal area  law;
complete non-Abelian expressions are therefore given only where useful, e.g.,
to clarify a point.  Our method consists in expressing the usual canonical
matrix elements from (\ref{lagrangian}) in terms of Wilson loop expectation
values  derived from the MA law. In the effective theory these become
first-quantized operators acting on state vectors. The results are fully
relativistic.

\section{Wilson loop expectation values }

We begin with the Lagrangian density assuming that the MA  analysis has been
carried out; hence fermion and gauge fields are interrelated at the outset,
effectively reducing the degrees of freedom in (\ref{lagrangian}). The idea is
simple: gauge fields specified by MA relations are taken to be "external" to
the Lagrangian; Euler-Lagrange non-linearities (\ref{el1}-\ref{el2})
 are thereby hidden within undetermined matrix elements.

 From the energy-momentum tensor of (\ref{lagrangian}) the conserved quantities
are the usual
\bea
H &=& \sum_j \int d^3 x \, \bar{q}_j [ \bgamma \cdot (- \imath \bnabla -g \bA)
+ \gamma^{0} gA_{0} +m_{j}] q_j  \ ,
\label{con}\\
\bP &=& \sum_j \int d^3 x \, q^{\dag}_j (- \imath \bnabla ) q_j \ ,
 \\
\bJ  &=& \sum_j \int d^3 x \, q^{\dag}_j [ \bx \times (- \imath \bnabla ) +
\frac{1}{2} \bsigma ] q_j \ ,
\label{con2}
\eea
which with the MA law,
\beq
\imath \ln \langle W({\cal C}) \rangle = a S_{min} \label{min}
\ ,
\eeq
define our bound state problem. Eventually, we take independent variations of
(\ref{min}) along the paths ${\cal C} $.

We consider the Wilson loop of a straight slice out of the $q\bar{q}$ minimal
world sheet from time $\tau^{\prime}$,
\bea
\langle W({\cal C}) \rangle &=& \frac{1}{3} \langle  \,{\rm Tr\, P
\,exp}(\imath g \oint_{\cal C} dx^{\mu} A_{\mu}(x)) \rangle  \nonumber
 \\
&=& \frac{1}{3} \langle  \,{\rm Tr\, P \,exp} ( -\imath g
\int_{\tau_1^{\prime}}^{\tau_{1f}} d\tau \,[A_0(z_1) - \dot{\bz}_1 \cdot
\bA(z_1)]) \, U(z^{\prime}_2,z^{\prime}_1) \nonumber \\
&\times&  {\rm P \, exp}( -\imath g \int_{\tau_{2f}}^{\tau_2^{\prime}} d\tau
\,[A_0(z_2) - \dot{\bz}_2 \cdot \bA(z_2)] ) \,  U(z_{1f},z_{2f})\, \rangle\ .
\nonumber \label{loop}
\eea
The $U$'s here are straight-line path ordered exponentials, and the average is
taken over gauge fields,
\beq
\langle \vartheta \rangle = \frac{ \int {\cal D} A  \exp(\imath S[A] )
\vartheta [A]}{\int {\cal D} A  \exp(\imath S[A] ) }\ ,
\eeq
where the pure gauge action  $ S[A] $ includes a gauge fixing term.

The c-number path variables for a given gauge field ($ \bz_i $  and  $
\dot{\bz}_i $) are related to coordinate and mechanical-momentum matrix
elements,
\bea
\bz_i &\equiv&  \bz_i(\tau ; A) = \langle q_i(\tau ; A) | \; \hat{\bx} \; |
q_i(\tau ; A) \rangle \ ,  \label{vel} \\
m_i \gamma_i \, \dot{\bz}_i &\equiv&  \langle q_i(\tau ; A) | \; ( \hat{\bp}
-g\bA(\hat{x})) \;  | q_i(\tau ; A) \rangle  \, ,\\
 \gamma_i &\equiv& \frac{1}{\sqrt{1- \dot{\bz}^2_i}}\ .
\eea
where  $ q_i $ is either a particle or antiparticle Euler-Lagrange solution
(\ref{el1}-\ref{el2}) with
a classical gauge field. These valence or quenched solutions prevent the
possibility of time-backtracking \cite{dubin} not present in the MA picture. If
$\chi$ is the meson wavefunction
in a product space of $|q_{1}\ran $ and $|q_{2}\ran $,
a  reasonable ansatz is the auxiliary condition
\beq
\chi \rar  \Lambda_{+}^{(1)}\chi \Lambda_{-}^{(2)}
- \Lambda_{-}^{(1)}\chi \Lambda_{+}^{(2)}\ ,
\eeq
where $ \Lambda_{\pm}^{(i)}$ are positive or negative energy
 projection operators. In this approximation, the Casimir operators reappear in
the Hamiltonian (\ref{con}).

On the contrary, we point out that the quark degrees of freedom have not yet
entered the dynamics except as a means to specify the physical gauge field
configurations. The above identifications are made for later use.

The enclosed area we parameterize  in the Nambu-Goto form,
\bea
S &=& a \int^{\tau_f}_{\tau^{\prime}} d\tau \int_0^1 d\sigma[
-\dot{x}^2 x'^2 +(\dot{x}\cdot x')^2]^{1/2}  \nonumber \\
&=& a \int^{\tau_f}_{\tau^{\prime}} d\tau \int_0^1 d\sigma \, {\cal S}
\eea
with $ x_{\mu} = x_{\mu}(\tau,\sigma), \, x_0 = \tau $,
$\dot{x}_{\mu}=\frac{\p x_{\mu}}{\p \tau}$, and
$x'_{\mu}=\frac{\p x_{\mu}}{\p \sigma}$.
 At the boundary, $ \bx(\tau,1) = \langle \bz_1(\tau) \rangle $ and $
\bx(\tau,0) = \langle \bz_2(\tau) \rangle $. In the usual straight-line and
equal-times approximations, which we assume here, the minimum is given by
\bea
\bx_{min}(\tau,\sigma) &=& \sigma \langle \bz_1(\tau) \rangle
+(1-\sigma)\langle \bz_2(\tau) \rangle\ .
\eea
A general path variation of (\ref{min}) for fixed endpoints and fixed time,
\bea
\imath \frac{\delta \langle W({\cal C}) \rangle }{\langle W({\cal C}) \rangle }
&=& \frac{-\imath }{\langle W({\cal C}) \rangle}
\cdot \imath \int_{\tau^{\prime}}^{\tau_f} d\tau \frac{1}{3} \bigg\langle
\,{\rm Tr\, P }  \exp{(\imath g \oint_{\cal C} dx^{\mu} A_{\mu}(x))}
 \nonumber
\\  &\times&   g \bigg\{ (\delta \bz_1  \cdot \frac{\p}{\p \bz_1} + \delta
\dot{\bz}_1 \cdot \frac{\p}{\p \dot{\bz}_1} ) [A_0(z_1) - \dot{\bz}_1 \cdot
\bA(z_1)] \nonumber \\
 &-& (\delta \bz_2  \cdot \frac{\p}{\p \bz_2} + \delta \dot{\bz}_2 \cdot
\frac{\p}{\p \dot{\bz}_2} ) [A_0(z_2) - \dot{\bz}_{2} \cdot \bA(z_2)] \bigg\}
\bigg\rangle  \nonumber \\
&=& a \int^{\tau_f}_{\tau^{\prime}} d\tau \int_0^1 d\sigma \left[\frac{\p {\cal
S}}{\p x'^{\mu}}   \delta x'^{\mu} + \frac{\p {\cal S}}{\p
\dot{x}^{\mu}}  \delta \dot{x}^{\mu} \right]_{x_{min}} \ , \label{variation}
\eea
specified to $ \delta \bz_1 = \delta \bz_2 $,
obtains in the center of momentum (see Appendix for details)
\bea
\langle \langle g\bA(z) \rangle \rangle &\equiv& \frac{1}{\langle W({\cal C})
\rangle } \frac{1}{3} \bigg\langle  \,{\rm Tr\, P } {\rm exp}(\imath g
\oint_{\cal C} dx^{\mu} A_{\mu}(x)) g \bA(z)  \bigg\rangle  \nonumber \\
&=& a \int_0^{\langle z \rangle } dx \,  \gamma_{\perp} \dot{\bx}_{\perp}
\equiv \bp_{t}\ ,\label{vec}
\eea
which are the desired Wilson loop expectation values with reference points
chosen at the origin, $ \bA(0)=0 $, and "$ \perp $" defined relative to the
straight line, $ \bv_{\perp} \equiv \bv - (\bv \cdot \hat{\br}) \hat{\br} $. It
should be clear that the gauge fields quantized in this average carry spatial
dependence in the radial coordinate only. Evidently, this proceedure selects
the physically realizable transverse polarizations.
The time component is obtained by simple differentiation of (\ref{min}) with
respect to $ \tau$,
\bea
\frac{d}{d\tau} \; \imath \ln \langle W({\cal C})\rangle &=& \frac{\imath
}{\langle W({\cal C}) \rangle }
 \imath  \frac{1}{3} \bigg\langle  \,{\rm Tr\, P }  \exp{(\imath g \oint_{\cal
C} dx^{\mu} A_{\mu}(x))}  \nonumber \\  &\times&   g \{ A_0(z_1) - A_0(z_2)
\nonumber  - [\dot{ \bz}_{1} \cdot \bA(z_1) -\dot{ \bz}_{2} \cdot \bA(z_2) ] \}
 \bigg\rangle  \nonumber \\
&=&  - \, a  \int_0^1 d\sigma\; {\cal S}_{x_{min}}\ ,
\label{a0}
\eea
yielding (using the
proceedure described in the Appendix)
\beq
\langle \langle g A_0(z) \rangle \rangle = a \int_0^{\langle z \rangle } dx \,
\gamma_{\perp}\equiv H_{t} \label{ao}\ ,
\label{a00}
\eeq
and also
\bea
\langle \langle \bz \times g\bA(z) \rangle \rangle
&=& a \int_0^{\langle z \rangle } dx ( \bx \times \gamma_{\perp}
\dot{\bx}_{\perp} )\equiv \bL_{t} \, .
\label{za}
\eea

\section{Relation to Flux Tube Model}
With the  Wilson loop expectation values we
re-express conserved quantities (\ref{con}-\ref{con2}) in terms of coordinate
and velocity matrix elements. In the  effective theory these are promoted to
noncommuting quantum operators, observables requiring symmetrization. We give
the final expressions
\bea
H &=& \sum_{i=1}^2  \, \balpha_i \cdot \left[ m_i \gamma_i \, \dot{\bx}_i
\right]_{sym}  + \quad \Bigl[ \langle \langle g A_{0}(x_i) \rangle \rangle
\Bigr]_{sym}  + \beta m_i \ ,
\label{flux1}\\
\bP &=& \sum_{i=1}^2 \quad \left[  m_i \gamma_i \, \dot{\bx}_i \right]_{sym} +
\quad \Bigl[ \langle \langle g \bA(x_i) \rangle \rangle \Bigr]_{sym} \quad = \;
0 \label{flux}\ , \\
\bJ &=&  \sum_{i=1}^2 \quad \left[ \bx_i \times  m_i \gamma_i \, \dot{\bx}_i
\right]_{sym} + \quad \Bigl[ \langle \langle \bx_i \times g \bA(x_i) \rangle
\rangle \Bigr]_{sym} + \frac{\bsigma_i}{2} \quad .
\label{flux2}
\eea
This compares well with other results from the Wilson loop. Both spin and
spin-independent Hamiltonians of \cite{eichten}, for example, are reproduced on
semi-relativistic reduction of (\ref{flux}).
Equations (\ref{flux1}-\ref{flux2}) in fact define the relativistic flux tube
model \cite{Isgur} as formulated by the present authors in \cite{low,ovw}.
There, dynamical relations are derived from a heuristic Lagrangian in which the
flux tube is fashioned as a simple constant energy density. The "tube
operators" are equivalent to the Wilson loop expectation values above.
This straightforward derivation serves to clarify the model's
close relation to the underlying QCD.

\section{Conclusion}

We have demonstrated that under the same assumptions as in previous work
\cite{eichten,dubin}
that the Wilson loop area law and the QCD Lagrangian yields a
relativistically
valid picture of dynamical confinement. We emphasize that we have not
``solved'' QCD, but that our ignorance can be distilled into three
loop expectation values each of which
is equal to that of the mechanical RFT model. We find that
\bea
\lan\lan g A_{0}\ran\ran &=& H_{t}\ , \\
\lan\lan g \bf{A}\ran\ran &=& {\bf p}_{t}\ ,\\
\lan\lan {\bf r}\times g{\bf A}\ran\ran &=& {\bf L}_{t}\ ,
\eea
for the tube energy, momentum, and angular momentum respectively.
In addition we
have shown that the natural
equation
of motion is the Salpeter equation in which the ``covariant
tube substitution'' \cite{ovw} has been made.

The verification
of the RFT model structure promises more  than to legitimize
a physically reasonable model. The close relation of the RFT model
to lattice QCD should shed light on both subjects. In addition, a
systematic program of improving the RFT model to include field
fluctuations \cite{Isgur} can now be envisioned.

\appendix

\section{Appendix}

Equation (\ref{vec}) follows from the physical gauge field's spatial dependence
in the straight line and equal times approximations
\bea
A(\br) &\sim & \sum_{\bk} \exp(-\imath \bk \cdot \br ) \, \to \, \sum_k
\exp(-\imath k r ) \ .
\eea
Then small angular variations of a given path at fixed time leave $ A $
unchanged. It will suffice to consider the transverse part of (\ref{variation})
with
\bea
\delta \bz_i(\tau;A) &=& \delta \bz_i(\tau;A^{\prime}) \, \equiv \, \delta
\bz_i \ ,
\eea
for all gauge fields at a given $ \tau $. Naturally,
\bea
\delta \langle \bz_i \rangle &=& \delta \bz_i \, .
\eea
We write the Wilson loop in discrete form,
\bea
W({\cal C}) &=& \frac{1}{3} {\rm Tr} {\rm \ P} \prod_{n=0}^N \exp( \imath g
\Delta t_n [ A_{0}(z_{1n})-\dot{\bz}_{1n} \cdot \bA(z_{1n}) ] )
U(z_2^{\prime},z_1^{\prime})\nonumber
 \\
&\times&  \exp( -\imath g \Delta t_n [ A_{0}(z_{2n})-\dot{\bz}_{2n} \cdot
\bA(z_{2n}) ] ) U(z_{1f},z_{2f}) \ ,
\eea
and take the path variations of (\ref{min}) according to
\bea
\delta (\imath \ln \langle W({\cal C})\rangle ) &=& \frac{\imath}{\langle
W({\cal C})\rangle} \delta \langle W({\cal C})\rangle \nonumber
\\ &=& \frac{\imath}{\langle W({\cal C})\rangle} \bigg\langle \sum_{n=0}^{N}
(\delta \bz_{1n} \cdot \frac{\p}{\p \bz}_{1n} + \delta \dot{\bz}_{1n} \cdot
\frac{\p}{\p \dot{\bz}}_{1n}
\nonumber \\
&+& \, \delta \bz_{2n} \cdot \frac{\p}{\p \bz}_{2n} + \delta \dot{\bz}_{2n}
\cdot \frac{\p}{\p \dot{\bz}}_{2n}) W({\cal C})\bigg\rangle \ ,
\eea
giving equation (\ref{variation}),
\bea
&& \frac{1}{\langle W({\cal C}) \rangle}
\int_{\tau^{\prime}}^{\tau_f} d\tau \frac{1}{3} \bigg\langle  \,{\rm Tr\, P }
\exp{(\imath g \oint_{\cal C} dx^{\mu} A_{\mu}(x))} \sum_{i=1}^{2}
 (\delta \bz_i  \cdot \frac{\p}{\p \bz_i} + \delta \dot{\bz}_i \cdot
\frac{\p}{\p \dot{\bz}_i} ) \, f(z_i,\dot{\bz}_i) \bigg\rangle  \nonumber \\
& & \qquad \qquad \qquad
 = \  \ a \int^{\tau_f}_{\tau^{\prime}} d\tau \int_0^1 d\sigma \left[\frac{\p
{\cal S}}{\p x'^{\mu} } \frac{\p }{\p \sigma} \delta x^{\mu} + \frac{\p {\cal
S}}{\p
\dot{x}^{\mu} }  \frac{\p }{\p \tau} \delta x^{\mu} \right]_{x_{min}} \ ,
\eea
where
\beq
f(z_i,\dot{\bz}_i) = (-1)^i [ \dot{\bz}_i \cdot g\bA(z_i) - gA_0(z_i)]\ .
\eeq
The temporal derivative is transferred from the variations by partial
integration, so that
\bea
 &\frac{1}{\langle W({\cal C}) \rangle}
\int_{\tau^{\prime}}^{\tau_f} d\tau \sum_i \frac{1}{3} \bigg\langle  \,{\rm
Tr\, P } \exp{(\imath g \oint_{\cal C} dx^{\mu} A_{\mu}(x))} ( -
\frac{d}{d\tau} \frac{\p}{\p \dot{\bz}_i} + \frac{\p}{\p \bz_i}  ) \,
f(z_i,\dot{\bz}_i) \bigg\rangle \cdot \delta \bz_i \hspace*{+1cm}& \nonumber \\
&=  a \int^{\tau_f}_{\tau^{\prime}} d\tau \int_0^1 d\sigma \Bigg[-
\left(\frac{\p {\cal S}}{\p \bx' } \right)_{x_{min}} \hspace*{-5mm}\cdot
(\delta \bz_1 - \delta \bz_2 )
+ \frac{d}{d\tau} \left( \frac{\p {\cal S}}{\p \dot{\bx} } \right)_{x_{min}}
\hspace{-5mm}\cdot  [ \, \sigma \delta \bz_1 + (1-\sigma) \delta \bz_2 \,]
\Bigg] \ ,&
\eea
yielding in the $ \delta \bz_1 = \delta \bz_2 \, ( \equiv \delta \bz_{\perp} )
$ case of interest
\bea
- \sum_i \langle \langle \frac{\p}{\p \dot{\bz}_i} f(z_i,\dot{\bz}_i) \rangle
\rangle &=&
a \int_0^1 d\sigma \left( \frac{\p {\cal S}}{\p \dot{\bx} } \right)_{x_{min}}\
,
\eea
or
\beq
\langle \langle g\bA(z_1) - g\bA(z_2) \rangle \rangle = a | \langle \bz_1
\rangle - \langle \bz_2 \rangle | \int_0^1 d\sigma \frac{\sigma \langle
\dot{\bz}_{1\perp} \rangle +(1-\sigma) \langle \dot{\bz}_{2\perp} \rangle }{
[1- (\sigma \langle \dot{\bz}_{1\perp} \rangle +(1-\sigma) \langle
\dot{\bz}_{2\perp} \rangle )^2 ]^{1/2}}\  .
\eeq
This obtains (\ref{vec}) when reference points are chosen at the origin of
coordinates.

Also, with the $A_{0}$ reference point chosen at the
origin, $A_{0}(0) = 0$, and (\ref{a0}) can be written
\bea
&\frac{1}{\lan W(C) \ran} \frac{1}{3} \frac{
{\rm Tr\ P\ }\int d A_{t}g [A_{t}(z_{A_{t}})
-\dot{\bz}_{A_{t}}\cdot \bA_{t}(z_{A_{t}})]
\int {\cal D}A_{\not{=}t}\exp{(ig\oint_{C} dx^{\mu} A_{\mu})}
\exp{(iS[A])} }{\int {\cal D} A  \exp(\imath S[A] ) }&
\nonumber \\
&= a \lan z_{t}\ran \int_{0}^{1}d\sigma
 [1-\sigma^2\lan \dot{\bz}_{t}\ran^2_{\perp}]^{1/2}\ ,&
\label{a0ap}
\eea
with
\beq
\lan \dot{\bz}_{t}\ran = \frac{\int dA_{t} \dot{\bz}_{A_{t}}
 \int {\cal D}A_{\not{=}t}\exp{(iS[A])} }
{\int {\cal D} A  \exp(\imath S[A] ) }
\ .
\eeq
Then, the functional derivative of (\ref{a0ap}) with respect
to $\dot{\bz}_{A'_{t}}$ is
\bea
&\frac{1}{\lan W(C) \ran} \frac{1}{3} \frac{
{\rm Tr\ P\ }(-g\bA'_{t})
\int {\cal D}A_{\not{=}t}\exp{(ig\oint_{C} dx^{\mu} A_{\mu})}
\exp{(iS[A])} }{\int {\cal D} A  \exp(\imath S[A] ) }&
\nonumber \\
&= a \lan z_{t}\ran \int_{0}^{1}d\sigma (-\sigma^{2})
\lan \dot{\bz}_{t}\ran_{\perp}
 [1-\sigma^2\lan \dot{\bz}_{t}\ran^2_{\perp}]^{-1/2}
\frac{\int{\cal D}A_{\not{=}t} \exp{(iS[A])} }
{\int {\cal D} A  \exp(\imath S[A] ) }
\ .&\label{a0ap2}
\eea
Taking the scalar product of the above with
$\dot{\bz}_{A'_{t}}$, and integrating
over $A'_{t}$, gives
\beq
\langle \langle \bz \cdot g\bA \rangle \rangle = a \int_0^{\langle z \rangle }
dx \,
 \dot{\bx}_{\perp}^{2}\gamma_{\perp} \ ,
\eeq
which is used in obtaining (\ref{a00}).
Taking the vector product of (\ref{a0ap2})
with $\dot{\bz}_{A'_{t}}$ yields (\ref{za})
after integration over $A'_{t}$,
\bea
\langle \langle \bz \times g\bA(z) \rangle \rangle
&=& a \int_0^{\langle z \rangle } dx ( \bx \times \gamma_{\perp}
\dot{\bx}_{\perp} ) \, .
\eea

\vskip 1cm
\begin{center}
ACKNOWLEDGMENTS
\end{center}
We would like to thank N. Brambilla and G. Prosperi for reading
the manuscript and making many helpful comments.
This work was supported in part by the U.S. Department of Energy
under Contract Nos.  DE-FG02-95ER40896 and DE-AC05-84ER40150,
the National Science
Foundation under Grant No. HRD9154080,
and in part by the University
of Wisconsin Research Committee with funds granted by the Wisconsin Alumni
Research Foundation.

\end{document}